\documentclass[useAMS,usenatbib]{mn2e}
\usepackage{graphicx,graphics,amsmath,amssymb,psfig}
\title[Detectability of extragalactic fast transients]
{On the detectability of extragalactic fast radio transients}
\author[D.~R.~Lorimer et al.]
{D.~R.~Lorimer$^{1,2,3}$,
A.~Karastergiou$^{3}$, M.~A. McLaughlin$^{1,3}$ and S.~Johnston$^{4}$\\
$^1$ Department of Physics, West Virginia University, PO~Box~6315, Morgantown,
WV~26506, USA\\
$^2$ National Radio Astronomy Observatory, PO~Box~2, Green Bank, WV~24944, USA\\
$^3$ Astrophysics, University of Oxford, Denys Wilkinson Building, Keble Road, Oxford OX1 3RH\\
$^4$ CSIRO Astronomy and Space Science, Australia Telecope National Facility,
PO Box 76, Epping, NSW 1710, Australia}
\begin{document}
\maketitle
\begin{abstract}
Recent discoveries of highly dispersed millisecond radio bursts by
Thornton et al. in a survey with the Parkes radio telescope at 1.4~GHz
point towards an emerging population of sources at cosmological
distances whose origin is currently unclear. Here we demonstrate that
the scattering effects at lower radio frequencies are less than
previously thought, and that the bursts could be detectable at
redshifts out to about $z=0.5$ in surveys below 1~GHz. Using a source
model in which the bursts are standard candles with bolometric
luminosities $\sim 8 \times 10^{44}$~ergs/s uniformly distributed per unit
comoving volume, we derive an expression for the { observed} peak flux 
{ density} as a
function of redshift and use this, together with the rate estimates
found by Thornton et al. to find an empirical relationship between
event rate and redshift probed by a given survey.  The non-detection
of any such events in Arecibo 1.4~GHz survey data by Deneva et al.,
and the Allen Telescope Array survey by Simeon et al. is consistent
with our model. Ongoing surveys in the 1--2~GHz band should result in
further discoveries. At lower frequencies, assuming a typical
{ radio} spectral index $\alpha=-1.4$, the predicted peak flux
densities are 10s of Jy. As a result, surveys of such a population
with current facilities would not necessarily be sensitivity limited
and could be carried out with small arrays to maximize the sky
coverage. We predict that sources may already be present in 350-MHz
surveys with the Green Bank Telescope. Surveys at 150~MHz
with 30~deg$^2$ fields of view could detect one source per hour above
30~Jy.
\end{abstract}

\begin{keywords}
surveys: radio --- scattering --- intergalactic medium
\end{keywords}

\section{Introduction}

In the last few years, a small population of sources emitting
short-duration (``fast'') transient radio bursts has been found. The
bursts are bright, last no more than a few milliseconds, are broadband
and show the characteristic signature of dispersion by a cold plasma
medium. Dispersion results in a frequency-dependent arrival time
across the band, proportional to the integral of the electron column
density along the line of sight, otherwise known as the dispersion
measure (DM). The prototypical fast radio burst (FRB) was discovered
by \citet{lbm+07} at a DM of 375~cm$^{-3}$~pc in archival pulsar
survey data of the Magellanic clouds \citep{mfl+06}.  In a reanalysis
of Parkes Multibeam Pulsar Survey data, \citet{kkl+11,kskl12}
identified an FRB with a DM of 746~cm$^{-3}$~pc. Recently,
\citet{tsb+13} report the discovery of four further FRBs at DMs in the
range 550--1100~cm$^{-3}$~pc.  Given detailed models of the Galactic
electron column density distribution, these high DMs place these
sources far beyond the extent of the Galaxy and signify the emergence
of a population of cosmological transients with exciting applications
as probes of new physics and the intergalactic ionized medium.

Dispersion is not the only frequency-dependent effect incurred from
propagation through an ionized plasma. FRBs will also be scattered due
to inhomogeneities in this medium, with scattered rays arriving at the
telescope later than those traveling on the direct line of sight
\citep[e.g.,][]{ric90}.  The resulting observed scattered pulse can be
well approximated in most cases by a Gaussian pulse convolved with an
exponential tail, with temporal constant $\tau_{sc}$ scaling with
frequency $\nu$ as $\tau_{sc}\propto \nu^{\eta}$. Although it was
often assumed that $\eta=-4.4$, as expected for Kolmogorov turbulence
\citep{ric90}, \citet{lkm+01} found a flatter dependence
where $\eta=-3.4 \pm 0.1$.  Later, from a larger sample of pulsars,
\citet{bcc+04} found that $\eta=-3.9 \pm 0.2$ and presented an
empirical relation between $\tau_{sc}$ and DM (see \S 2). These two 
quantities correlate well for the Galactic population of pulsars, 
albeit with a dispersion of up to 
an order of magnitude on either side of the DM-$\tau_{sc}$ curve.

Using the results of surveys at 1400~MHz, we investigate event rate
predictions for surveys at 150~MHz and 350~MHz. The 150~MHz band is
pertinent given the advent of LOFAR \citep{sha+11} as a wide-field of
view low-frequency transient monitor. The 350~MHz band is covered by
all-sky pulsar surveys being carried out at the Green Bank Telescope
\citep{blr+13,lbr+13,rsm+13} and at Arecibo \citep{dsm+13}.
In \S 2 we demonstrate that the scattering in
low-frequency surveys is less severe if FRBs are at cosmological
distances. In \S 3 we describe a simple model that provides testable
event rate predictions for ongoing and future surveys. We discuss
these results in \S 4 and present our conclusions in \S 5.

\section{Anomalous scattering in extragalactic fast transients}
\label{sec:ascat}

\begin{figure} 
\includegraphics[width=0.49\textwidth]{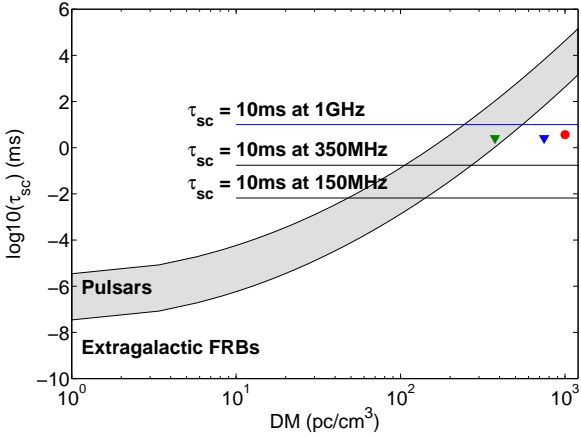}
\caption{Scattering time at 1~GHz versus dispersion measure showing
  the radio pulsars (adapted from Bhat et al.~2004) along with current
  scattering constraints on FRBs. The green and blue triangles
  indicate the scattering timescale upper limits of 1~ms at 1.4 GHz
  for the FRBs discussed in Lorimer et al. (2007) and Keane et
  al. (2012), scaled to 1~GHz. The red circle indicates the scattering
  timescale of 1~ms scaled to 1~GHz measured for one of the FRBs
  discussed in Thornton et al. (2013). The other FRBs, with DMs of
  944, 723, and 553~cm$^{-3}$~pc have scattering timescale upper 
  limits of 1~ms.}
\label{fg:bhat}
\end{figure}

\citet{lbm+07} discussed the possibility of detecting FRBs with
low-frequency telescopes. The burst they detected showed evolution of
pulse width with frequency, but they could not determine whether this
was intrinsic or due to scattering. The $\sim$ms scattering expected
at a DM of 375~cm$^{-3}$~pc at 1~GHz scales to a scattering timescale
of order 1~s at 150~MHz, leading to their suggestion that the pulse
would be undetectable at those frequencies.  However, of the 6 FRBs
known so far, only one ({ FRB~110220}) 
has had a measurable scattering timescale at
1.4~GHz. The observation that the FRBs all lie significantly below the
bulk of pulsars in the Bhat et al.~(2004) DM-$\tau_{sc}$ curve,
suggests that, for a particular DM, there is less scattering for an
extragalactic source than would be expected from interpreting the curve.

For FRBs of extragalactic origin, the total scattering is made up of two
contributions: interstellar scattering in the host galaxy and our own
Galaxy, and intergalactic scattering caused by the intervening
intergalactic medium. Despite the fact that most of the scattering
material is found in the interstellar medium of the host or our
Galaxy, geometrical considerations \citep[e.g.,][]{wil72} suggest that
contributions to scattering from media near the source or near the
observer are expected to be small. To estimate the impact of this, we
rewrite equation (2) from \citet{mjsn98} in terms of the scattering
induced by a screen at a fractional distance $f$ along the line of
sight. As $f \rightarrow 0$, the screen is close the observer, and as
$f \rightarrow 1$, the screen is close to the source. In this case the
scattering time
\begin{equation}
\tau_{sc} = 4 \tau_{\rm max} (1-f) f,
\end{equation}
where $\tau_{\rm max}$ is the maximum scattering induced for a screen
placed midway along the line of sight ($f=0.5$).  For scattering
originating from a host galaxy at cosmological distances, $(1-f)$ is
the size of the host galaxy divided by the distance to the source,
i.e.~$(1-f) \simeq 50~{\rm kpc}/ 500~{\rm Mpc}= 10^{-4}$ for a Galaxy
with the extent of the Milky Way at the distance inferred for the
\citet{lbm+07} FRB.  Therefore, the scattering effects of our own
Galaxy or the host galaxy can be essentially neglected for bursts at
cosmological distances.

The same geometric considerations suggest that the most efficient
scattering along the line of sight towards FRBs of extragalactic origin will
occur in the medium near the midway point. Measurements of the
scattering measure (SM) along extragalactic lines of sight\footnote{SM
is the line integral of the electron density wavenumber spectral
coefficient $C_n^2$ \cite[for details, see, e.g.,][]{cr98}.}  by
\citet{lof+08} suggest values that are typically $\le 10^{-4}$ SM of
Galactic lines of sight, despite the large distances to extragalactic
sources.  Given that $\tau_{sc} \propto {\rm SM}^{6/5}$ and the
distance \citep{cl02}, similar pulse broadening times could be
observed for a Galactic source at 5 kpc as for an extragalactic source
at $\approx$300 Mpc. In Fig.~\ref{fg:bhat}, we show our adaptation of
the DM-$\tau_{sc}$ curve and plot the FRB scattering timescales scaled
to an observing frequency of 1~GHz. Here we see that the upper limit
on the scattering timescale for the FRB of \citet{kskl12} lies well
below the expected trend for Galactic pulsars, as does the timescale
for the one event of four for which \citet{tsb+13} were able to
measure $\tau_{sc}$. Taking the 1-ms of scattering at 1.4~GHz of this
event, which scales to 3.66~ms at 1~GHz, and comparing it to the
predicted value of $10^{3.63}$~ms, suggests that rescaling the
\citet{bcc+04} equation:
\begin{equation}
\log \tau_{sc} \simeq -6.5 + 0.15 (\log {\rm DM}) +  1.1 (\log {\rm DM})^2
-3.9 \log f
\end{equation}
so that the leading term is --9.5 rather than --6.5
provides an estimate of the expected scattering for
extragalactic sources. In this expression, DM takes its usual units
of cm$^{-3}$~pc and the frequency, $f$, is in GHz.
Note that, given that only one measurement of the
scattering timescale has been made so far, it is likely that this is
an upper limit to the average amount of scattering as a function of DM.

In summary, based on the theoretical expectations of low scattering
towards FRBs of extragalactic origin and the recent 1.4~GHz observations
in which little or no scattering is observed,
FRB scattering at frequencies below 1~GHz
is also expected to be substantially less than would be the case if
they were of Galactic origin. This raises the possibility that they
can be detectable in surveys at lower frequencies.  In the remainder
of this paper, we discuss the implications of this conclusion and make
predictions using a population model.

\section{Event rate predictions for fast transient surveys}

The events reported by \citet{tsb+13}, along with those previously
\citep{lbm+07,kskl12} imply a substantial population of transients
detectable by ongoing and planned radio surveys.  While an
investigation of event rates was recently carried out by
\citet{mac11}, the analysis assumed Euclidean geometry, ignoring
cosmological effects. The results of \citet{tsb+13}, where significant
redshifts ($z \sim 0.7$) are implied by the high dispersion measures,
imply that propagation effects in an expanding universe must be taken
into account. To begin to characterize this population and make
testable predictions, we consider { the simplest possible}
cosmological model in which
FRBs are standard candles with constant number density per unit
comoving volume. The former assumption is justified for source models
in which the energetics do not vary substantially. Along with the
latter assumption, as we demonstrate below, this model is eminently
testable by ongoing and future fast-sampling radio surveys.

\begin{figure*} 
\centerline{\psfig{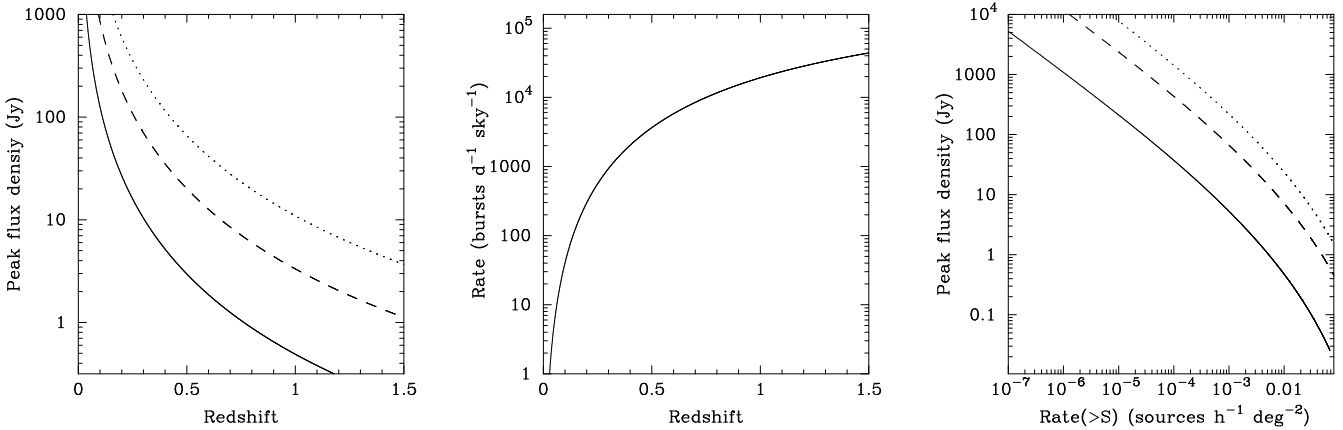}} 
\caption{Predictions from our FRB population model.
  The left panel shows flux--redshift 
  relationships for surveys carried out at
  1400~MHz (solid line), 350~MHz (dashed line) and 150~MHz (dotted
  line). The centre panel shows
  the event rate normalized such that at $z=0.75$ 
  the implied event rate is
  10,000 FRBs per day per sky as inferred by Thornton et
  al.~(2013). The right panel shows the predicted burst rates above some
  threshold flux density $S$ at 1400~MHz (solid line), 350~MHz (dashed
  line) and 150~MHz (dotted line).}
\label{fg:candles}
\end{figure*}

Considering the sources as standard candles implies that, for a given
survey at some frequency $\nu$, there is a unique correspondence
between the observed flux density and redshift probed. { To 
simplify matters, we consider a model in which
the pulses have a top-hat shape with some finite width. As
we show below, under these assumptions, the flux density--redshift
relationship is independent of pulse width.} To derive this
relationship, we use standard results \citep[e.g.,][]{hog99} and adopt
a flat universe where, for a source at redshift $z$, the comoving
distance
\begin{equation}
\label{eq:comoving}
D(z) =  \frac{c}{H_0} 
\int_0^z \frac{dz'}{\sqrt{\Omega_m(1+z')^3 + \Omega_{\Lambda}}}.
\end{equation}
Here $c$ is the speed of light, $H_0$ is the Hubble constant and the
dimensionless parameters $\Omega_m$ and $\Omega_{\Lambda}$ represent
the total energy densities of matter and dark energy respectively.
Following the latest results from {\it Planck} \citep{aaa+13} we adopt
a flat universe in which $H_0=68$~km~s$^{-1}$~Mpc$^{-1}$,
$\Omega_m=0.32$, and $\Omega_{\Lambda}=0.68$.  To obtain the peak flux
density $\bar{S}_{\rm peak}$ averaged over a certain bandwidth at some
frequency $\nu$, we model the energy released per unit frequency
interval in the rest frame, $E_{\nu'}$, using the power-law
relationship
\begin{equation}
  E_{\nu'} = k \nu'^{\alpha},
\end{equation}
where $k$ is a constant, $\nu'$ is the rest-frame frequency and
$\alpha$ is a spectral index. { Assuming a top-hat pulse
of width W' in the rest frame of the source,}
from conservation of energy, { for an observation over
some frequency}
band between $\nu_1$ and
$\nu_2$, we may write
\begin{equation}
\label{eq:speak0}
  \bar{S}_{\rm peak} 4 \pi D_L^2 (\nu_2-\nu_1) =
  \frac{\int_{\nu'_1}^{\nu'_2} E_{\nu'} d\nu'}{W'},
\end{equation}
{ where the luminosity distance $D_L=(1+z)D(z)$.}
{ It is worth noting here that the $(\nu_2-\nu_1)$
term on the left-hand side of this expression reflects the
fact that the pulse
has been obtained by dedispersion over this finite observing band. The
measured quantity, $\bar{S}_{\rm peak}$ is therefore
\begin{equation}
  \bar{S}_{\rm peak} = \frac{1}{\nu_2-\nu_1} \int_{\nu_1}^{\nu_2}
  S(\nu) d\nu ,
\end{equation}
where $S(\nu)$ represents the flux density per narrow frequency
channel within the band.
} { On the right hand side of equation \ref{eq:speak0},}
$\nu'_1=(1+z)\nu_1$ and $\nu'_2=(1+z)\nu_2$ are the lower and
upper extent of the observing band in the rest frame of the
source. Using these identities, and integrating equation
\ref{eq:speak0} for the case where $\alpha \ne -1$, we find
\begin{equation}
\label{eq:speak1}
  \bar{S}_{\rm peak}  =
  \frac{k (1+z)^{\alpha-1}
   (\nu^{\alpha+1}_2-\nu^{\alpha+1}_1)
   }{4 \pi D(z)^2 W' (\nu_2-\nu_1) (\alpha+1)}.
\end{equation}
To write equation \ref{eq:speak1} in terms of a luminosity model, we
note that the bolometric luminosity
\begin{equation}
  L = \frac{\int_0^\infty E_{\nu'} d\nu'}{W'} = \frac{k
  \left(
  \nu'^{\alpha+1}_{\rm high}-\nu'^{\alpha+1}_{\rm low}
  \right)
  }{W'(\alpha+1)},
\end{equation}
where the model
parameters $\nu'_{\rm low}$ and $\nu'_{\rm high}$ are respectively the
lowest and highest frequencies over which the source emits.  { Combining
these last two equations to eliminate $k/[W'(\alpha+1)]$, we find}
\begin{equation}
	\bar{S}_{\rm peak} = \frac{L (1+z)^{\alpha-1}}
{4 \pi D(z)^2 (\nu'^{\alpha+1}_{\rm high}-\nu'^{\alpha+1}_{\rm low})}
\left(
\frac{\nu_2^{\alpha+1}-\nu_1^{\alpha+1}}{\nu_2-\nu_1}
\right).
\end{equation}
To calibrate this flux---redshift relationship, based on the results
of \citet{tsb+13}, we adopt $\bar{S}_{\rm peak}=1$~Jy at $\nu = 1.4$~GHz at
$z=0.75$ and assume a spectral index $\alpha=-1.4$ which would be
appropriate if the emission process is { coherent, as observed
for the radio pulsar population \citep{blv13}}. With
this choice of parameters, and adopting $\nu'_{\rm low}=10$~MHz and
$\nu'_{\rm high}=10$~GHz, we require the bolometric luminosity { $L
\simeq 8 \times 10^{44}$~ergs/s}. 
Due to the normalization the exact choice of
$\nu'_{\rm low}$ and $\nu'_{\rm high}$ do not significantly affect the
rate calculations given below.  The results of this procedure, at
1400~MHz, 350~MHz and 150~MHz, with respective bandwidths of 350~MHz,
100~MHz and 50~MHz are shown in Fig.~\ref{fg:candles}.  { The
spectral indices of FRBs are currently not well constrained.
Thornton et al.~see no significant spectral
evolution within their 340~MHz bandwidth. If this turns out to
be the case over a broader frequency range, then
these extrapolated curves are overestimates of
the expected flux density and the 1400~MHz curves in Fig.~\ref{fg:candles}
would be more appropriate.}

Based on their results, \citet{tsb+13} compute an event rate, $R$, of
$10000^{+6000}_{-5000}$~bursts per day over the whole sky above 1~Jy
at 1400~MHz.  In our model, where all bursts are sampled out to a
redshift of 0.75, it is straightforward to scale this rate to other
redshifts via the ratio of the comoving volume $V(z)=(4/3)\pi D(z)^3$
enclosed compared to that at $z=0.75$. The rate--redshift relationship
is therefore $R(z) = R_{0.75} V(z)/ V_{0.75}$, where $R_{0.75}$ is the
Thornton et al.~rate at $z=0.75$, and $V_{0.75}$ is the comoving
volume out to $z=0.75$.  The results of this calculation are shown as
a function of $z$ in Fig.~\ref{fg:candles}. Also shown are
the predicted rates as a function of threshold flux density and
survey frequency.

\section{Discussion}

The model presented in the previous section and shown in
Fig.~\ref{fg:candles} makes a number of testable predictions for
ongoing and planned radio surveys.  Within the rate uncertainties
found by \citet{tsb+13}, assuming sensitivity out to some $z$, fitting
a cubic to the centre panel of
Fig.~\ref{fg:candles} provides the following good
approximation to the expected event rate out to $z=1$:
\begin{equation}
\label{eq:predictor}
R(<z) \simeq \left(\frac{z^2  + z^3}{4}\right) \,\, {\rm day}^{-1}\,{\rm deg}^{-2}.
\end{equation}

An important constraint for our model is that it should be consistent
with ongoing surveys at 1.4~GHz where assumptions about the source
spectral index are not required.  Currently, the most sensitive fast
transient search at 1.4~GHz is the ongoing Pulsar Arecibo L-band Feed
Array (PALFA) survey \citep{cfl+06,dcm+09} which has so far found no
FRBs.  Through detailed considerations, \citet{dcm+09} quantify the
effects of the greater depth probed by the PALFA survey and its
smaller field of view.  A comparison of the solid curve shown in the
right panel of
Fig.~\ref{fg:candles} of this paper and Fig.~8 from \citet{dcm+09}
shows that our model is currently not excluded by the lack of
detections in the PALFA survey. Our model is also consistent with the
``Fly's Eye'' survey carried out with the Allen Telescope Array (ATA)
by \citet{sbf+12} which was sensitive to pulses with peak fluxes
$>150$~Jy (assuming a 3~ms pulse width).  Our predicted rate of a few
times $10^{-6}$~FRBs~hr$^{-1}$~deg$^{-2}$ is below the ATA upper limit
of $2 \times 10^{-5}$~FRBs~hr$^{-1}$~deg$^{-2}$.

Further bursts are expected in the other 1.4~GHz multibeam surveys.
In the Parkes multibeam pulsar survey \citep{mlc+01}, where only one
FRB candidate has so far been found \citep{kskl12}, a simple scaling
of the Thornton et al.~(2013) rate leads to around 5--15 bursts in the
existing data. While a recent re-analysis by \citet{bnm12} did not find
any candidates in the DM range 200--2000~cm$^{-3}$~pc in addition to
the Keane et al.~event, the survey coverage along the Galactic plane
may imply that searches covering higher DM ranges are necessary.  A
search of archival data from the \citet{ebvb01} and \cite{jbo+09}
intermediate and high latitude surveys by \citet{bb10} revealed a
number of rotating radio transients, but no new FRBs. However, the DM
range covered in this effort (0--600~cm$^{-3}$~pc) was likely not
sufficient to sample a significant volume, based on the results
presented here. We suggest that reanalyses of these data sets may
result in further FRB discoveries.  We note also that the substantial
amount of time spent and field of view covered by the HTRU-North
survey at Effelsberg \citep{bck+13} mean around 20 FRBs are expected.

At 350~MHz, a significant fraction of the transient sky is currently
being covered by pulsar searches with the GBT. The drift-scan survey
carried out during summer 2007 \citep{blr+13,lbr+13,rsm+13}, acquired
a total of 1491 hours of observations and has so far discovered 35
pulsars. Based on the survey parameters given by \citet{lbr+13}, we
estimate the instantaneous field of view to be about 0.3~deg$^2$ and
the 10--$\sigma$ sensitivity for pulses of width 3~ms to be about
35~mJy. As can be inferred from Fig.~\ref{fg:candles}, a source at
this frequency is predicted to have a peak flux well in excess of this
threshold even out to $z>1$. As shown in Fig.~1, the expected
scattering of an FRB at 350~MHz should be below 10~ms for a DM of a
few hundred cm$^{-3}$~pc. Adopting a DM limit of 500~cm$^{-3}$~pc, and
assuming about 20\% of this DM is accounted for by the host galaxy and
the Milky Way, from the approximate intergalactic medium scaling law
\cite[where ${\rm DM} \sim 1200 z$~cm$^{-3}$~pc] []{iok03,ino04} we
expect a redshift limit of 0.33. From equation \ref{eq:predictor} we
infer a rate of about $4 \times 10^{-4}$ bursts per 0.3~deg$^2$ per
hour, or of order one FRB in the entire survey. Since the GBT survey
is not sensitivity limited and the steep gradient seen in the centre
panel of Fig.~\ref{fg:candles}, this prediction is subject to considerable
uncertainty. A further more sensitive GBT search is now underway with
the aim of covering the entire GBT sky (Stovall et al., in
preparation). A comprehensive analysis of these data, and ongoing
Arecibo 327~MHz surveys \citep{dsm+13} should provide
interesting constraints on the FRB population. If the same DM limit
can be reached by 150~MHz surveys with large fields of view, the
expected detection rate above 30~Jy with DMs below 500~cm$^{-3}$~pc is
$\sim 1$ event per day per 30 square degrees.

\section{Conclusions}

We have used the results of \citet{tsb+13} to calibrate a cosmological
model which predicts the rate of FRBs as a function of redshift. Our
assumption of uniform source density with comoving volume implies a
significant population of bursts detectable at moderate to low
redshifts by low-frequency ($\nu<1$~GHz) surveys. Moreover, our
assumption that the bursts are standard candles implies that such
events should be bright (10s of Jy or more) and would be readily
detectable by instruments with modest collecting areas. { Both of
these assumptions can be tested by ongoing and future surveys.}

An important
conclusion from this work is that low-frequency surveys should sample
as large a DM range as possible, and search for pulses over a wide
range of widths. An additional simplification we have made is that the
source spectra follow a power law with slope of --1.4. The event rates
predicted here will differ significantly if the spectrum deviates from
this form. Many of the current ongoing surveys and other large-scale
surveys expected over the next few years with LOFAR~\citep{sha+11}, 
{ the MWA} \citep{tgb+13} and the Australian
Square Kilometre Array Pathfinder \citep{mbb+10}{, and other facilities,}
will undoubtedly find
more FRBs and test the predictions presented here.

\section*{Acknowledgments}

This work made use of the SAO/NASA Astrophysics Data System.  DRL and
MAM acknowledge support from Oxford Astrophysics while on sabbatical
leave. We thank Sarah Burke-Spolaor, { Olaf Wucknitz}, 
{ and the referee, J-P Macquart, for useful comments on the manuscript.}

\end{document}